\documentclass[aps,prl,reprint,twocolumn,letter,superscriptaddress,nofootinbib]{revtex4-1}
\usepackage{amsmath, amssymb}
\usepackage{graphicx}
\usepackage{color}
\usepackage{ulem}
\usepackage{hyperref}

\hypersetup{
    pdfnewwindow=true,      
    colorlinks=true,       
    linkcolor=blue,          
    citecolor=blue,        
    filecolor=blue,      
    urlcolor=blue           
}

\begin{document}


\title{\large Neutron-Mirror-Neutron Oscillation and Neutron Star Cooling
}

\author{Itzhak Goldman}
\affiliation{Afeka College and Tel Aviv University, 6195001 Tel Aviv, Israel}

\author{Rabindra N. Mohapatra}
\affiliation{Maryland Center for Fundamental Physics and Department of Physics, University of Maryland, College Park, Maryland 20742, USA}

\author{Shmuel Nussinov}
\affiliation{Tel Aviv University, 6195001 Tel Aviv, Israel}

\author{Yongchao Zhang}
\affiliation{School of Physics, Southeast University, Nanjing 211189, China}

\begin{abstract}
It was pointed out in a recent paper that  the observed cooling rate of old, cold neutron stars (NS) can provide an upper limit on the transition rate of neutron to mirror neutron ($n-n'$). This limit is so stringent that it would preclude any discovery of {$n \to n'$ oscillation} in the current round of terrestrial searches for the process. Motivated by this crucially important conclusion, we critically analyze this suggestion and  note an interesting  new effect present in nearly exact mirror models for {$n \to n'$ oscillation}, which significantly affect this bound. The new element is the $\beta$ decay $n' \to  p'+ e' +\bar{\nu}'_{e}$, which creates a cloud of  mirror particles $n'$, $p'$, $e'$ and $D'$ inside the NS core. The $e'$ can ``rob'' the energy generated by the {$n \to n'$ transition} via   $e-e'$ scattering  enabled by the presence of   a (minute) milli-charge in mirror particles. This energy is emitted as unobserved mirror photons via fast mirror bremsstrahlung leading  to a relaxation  of this upper limit.

\end{abstract}

\date{\today}

\maketitle



{\bf Introduction:--}
Neutron stars  (NSs)  and their origin from Supernovae have played an important role in constraining physics beyond the standard model (BSM)~\cite{raffelt}. One class of BSM scenarios which can lead to new effects in NSs are the mirror models, which consist of a mirror sector coexisting with standard model (SM) and which contains a parity symmetric duplicate of the particles and forces of the SM~\cite{mirror}.  When the  mirror parity is nearly exact, all particles in the two sectors including the neutron and mirror neutron are nearly degenerate. This raises the possibility of neutrons oscillating to mirror neutrons ($n\to n'$)~\cite{bere1} if the sum of ordinary ($B$) and mirror ($B'$) baryon numbers is conserved. This phenomenon has been  proposed as a solution to the neutron lifetime anomaly~\cite{bere3}.  There are a number of experiments already carried out or planned to search for this {$n \to n'$ oscillation~\cite{expts}}. It is therefore important to know if there are any constraints on the $n-n'$ mixing parameter $\epsilon_{nn'}$ from astrophysical settings. Since  NSs are extremely rich in neutrons, they  are a perfect laboratory for testing  implications of {$n\to n'$ oscillation}.

{The transition of an ordinary neutron $n$ to a mirror neutron $n'$ is followed by a migration of the latter towards the NS center under gravity. The hole left will then be filled by another neutron at the Fermi level, and in the process energy is liberated~\cite{bere2}}. If the process is fast enough, it would lead to a fully mixed star. The resulting mass loss  of an NS
will not only lead to changes in the orbital period of a binary pulsar~\cite{GN}, but also affect the luminosity of {a single NS}~\cite{bere2, posp}.
The observational constraints on the rate of the binary periods for several binary pulsars were shown to lead to upper bounds on $\epsilon_{nn'}$ of  $10^{-13}$ eV~\cite{GMN}. On the other hand, taking the coldest NS, i.e. PSR J2144 -- 3933~\cite{Guillot:2019ugf}, it was argued in Refs.~\cite{bere2, posp} that one gets $\epsilon_{nn'}\leq 10^{-17}$ eV. Both the bounds are  valid for $n-n'$ mass difference up to 15 MeV~\cite{GMN}. This luminosity limit is particularly important, since currently planned terrestrial experiments are sensitive to $\epsilon_{nn'}$ at the level of $10^{-17} $ eV~\cite{expts}. Note that in terrestrial searches for $n-n'$ oscillation, to maintain coherent build-up of the mirror neutron wave function along the neutron beam,  and allow for such sensitive measurements, one must require a remarkably precise degeneracy between the neutron and its mirror partner of ${\delta_{nn'}}/m_n \leq 10^{-26}$ with $\delta_{nn'}\equiv{|m_{n'}-m_n|}$.

In this letter, we critically analyze the  luminosity bound, by following the evolution of the $n'$ generated in {$n \to n'$ transition} a bit longer. We observe that in almost exact mirror models, the  mirror neutrons generated inside the NS $\beta$ decay producing mirror fermions $e'$, $p'$ and $\bar{\nu}'_e$ leading eventually to a cloud of $e'$ and  deuterons $D'$. These mirror particles then provide a competing cooling channel via the emission of mirror photons $\gamma'$, and reduce the photonic signal claimed in Ref.~\cite{posp} considerably, relaxing the upper bounds on $\epsilon_{nn'}$.
For a relatively wide acceptable range of interactions between the ordinary and mirror sectors, mediated by the millicharge of mirror particles~\cite{holdom}, the  nucleons and electrons of the visible sector in this core region of the NS can transfer their energy to the mirror particles. The latter then emit this energy via mirror photons $\gamma'$, which do not interact with the ordinary nucleons and electrons and can freely escape. The philosophy of this paper is similar to that in Ref.~\cite{zurabb}.  The millicharge on mirror particles arises if  $\gamma$ and $\gamma'$ have kinetic mixing.

{\bf $n \to n'$ transition:--}
Initially, shortly after its birth, a NS is relatively hot and cools down via volume emission of neutrino pairs.
At the time of observation, the star may be still cooling off or, if some other sources of energy exist, it may have settled into a thermal steady state, with the thermal energy emitted
as electromagnetic radiation often as a black body radiation~\cite{Yakovlev1,Yakovlev2}. Let us apply this scenario to the pulsar PSR J2144 -- 3933. In a steady state, the NS black body luminosity  is given by the Stefan-Boltzmann formula ${\cal L}_{\rm NS} = 4\pi \sigma_{\rm SB} R^2 T_s^4$,
where $\sigma_{\rm SB}$ is the Stefan-Boltzmann constant, $R$ is the radius of the NS, and its external surface temperature $T_s$ is maintained by the constant internal energy source. If we have observational limits on the luminosity, this implies upper bounds on the rate of internal heat production.
It is important to note
that there is a $\sim$100 meter thick nuclear ``thermal blanket'' just under the surface~\cite{Yakovlev3}. It causes the internal temperature, which is almost uniform over the NS, to drop dramatically by a factor of  $\sim$100  as we move out from the inside across the blanket towards the surface. The estimated upper bound on surface temperature
$T_s\sim 42000$ K of the coldest pulsar PSR J2144 -- 3933 would  then correspond to the internal temperature $ T_{\rm int} \simeq 0.35$ keV, which would play an important role in obtaining upper bounds on any heat generating mechanism.


If the $n \to n'$ processes were the only source of heat supply, then in a steady state the overall $n-n'$ transition rate would be given by ${d{\cal N}_{n'}}/{dt} = {{\cal L}_{\rm NS}}/{\Delta E}$,
where $\Delta E\sim 30 $ MeV is the energy initially gained by ordinary nucleons in each $n\to n'$ transition. For PSR J2144 -- 3933, taking  $R=11$ km,
the rate of generating new mirror neutrons turns out to be:
\begin{eqnarray}
\label{eqn:dNpdt}
\frac{d {\cal N}_{n'}}{dt} \sim 0.45\times 10^{32} \left( \frac{T_s}{42000\ {\rm K}} \right)^4 \; {\rm sec}^{-1} \,.
\end{eqnarray}
During its long lifetime of 330 million years,  about
 $   {\cal N}_{n'} \sim 10^{48}$
neutrons would have converted into mirror neutrons. This comprises a tiny ${\cal N}_{n'}/{\cal N}_{n} \sim 10^{-9}$  fraction of the total neutron number ${\cal N}_n \sim 2 \times 10^{57}$ in the star, with no change of the gravity fields and of the local density profile of the ordinary NS.
Some pulsars have temperatures up to 100 times higher yielding $d{\cal N}_{n'}/d{t} \sim 10^{40}$ sec$^{-1}$, and were also used to bound high $\epsilon_{nn'}$ values~\cite{Yakovlev1, Yakovlev2}.


Neighboring neutrons rush into the ``hole'' formed by {$n \to n'$ transition}, and the work done in the process is $\sim 30$ MeV on average and becomes the kinetic energy of these nucleons. The nucleons collide with neighboring neutrons with density $ n_{N} \sim 10^{39}$ cm$^{-3}$, and very quickly settle into the spatially and temporally fixed internal temperature $T_{\rm int}$ ($ \sim  0.35 $ keV).
It should be noted that only the $f=kT/E_{F}$ fraction of nucleons and electrons in the high energy tail of the   degenerate Fermi-Dirac energy distribution are not Pauli blocked and can be excited (or de-excited) to higher (or lower) empty energy states, reducing the specific heat and the heat content $Q^*$ of the NS by a factor of $f$.  It is then given by
\begin{eqnarray}
\label{eqn:Qstar}
Q^*=  {\cal N}_n f^2 E_F  \,.
\end{eqnarray}
Upon using $kT \simeq 0.35$ keV for  PSR J2144 -- 3933 and $E_F = 30$ MeV,
we find $Q^* \sim   10^{52}$ keV,
with only the $f \sim 10^{-5}$ fraction of these end point ``active'' electrons partaking in electron scattering or any other dynamic processes, which will play an important role in the following calculations.



{\bf $n'$ decay and the $e'-D'$ fluid:--}
In connection with the extreme degeneracy of $n$ and $n'$, there are three extra light neutrinos and the mirror photon in exact mirror models. To bring about consistency between three extra neutrinos and an extra photon contributing to the energy density in the Big Bang Nucleosynthesis (BBN) epoch  of the universe  with the Planck data~\cite{Planck:2018nkj}, we require
that there be asymmetric inflation implemented~\cite{BDM}. This will remove the BBN problem by lowering the reheat temperature in the mirror sector by a factor of three, thus diluting the impact of the extra mirror neutrinos and the mirror photon on BBN. 

The $\beta$ decay $n'\to p' +e' +\bar{\nu}'_{e}$ of $n'$ proceeds in the same manner as $n$, and will have the the same rate of $\sim (800 \; \rm sec)^{-1}$ as $n$ decay in vacuum, so long as the Fermi energy of the electron is much smaller than the $Q$ value of 0.7 MeV of the $\beta$ decay.\footnote{As in the normal sector, there might be Zeeman effect in the mirror sector which lead to corrections to the $n'$ mass thus affecting the mirror $\beta$ decay. However, such effect is much leas important than the $n-n'$ mass splitting $\Delta M \sim {\cal O} ({\rm MeV})$ under gravity in the star, and can be safely neglected.} 
The $p'$s, like the $n'$s, are gravitationally bound to the NS, and local mirror charge neutrality forces {the number densities $n_{e',\,p'} (r)$ of $e'$ and $p'$ to be the same at all $r < R$, i.e. $n_{e'}(r) = n_{p'}(r)$}. The mirror neutrons and mirror protons slow down and form mirror deuterons $D'$,  since the process $p'+n'\to D'+\gamma'$
is faster than the inverse beta decay $ e'+p' \to n' +\nu'$.  All the $p'$s are ``eaten up'' to form $D'$, and the number of $e'$s that will remain is only half the number of $n'$ produced. The resulting $\gamma'$s escape taking away part of the energy released in $n \to n'$ transition but it does not drain the energy generated by neutrons falling from the Fermi surface, which is drained away via $e-e'$ scattering between the two sectors. Charge neutrality requires that $n_{e'}(r) = n_{D'}(r)$, {with $n_{D'}(r)$ the number density of $D'$}. The new processes we consider  are depicted schematically in Fig.~\ref{fig:1}.

\begin{figure*}[!t]
   \includegraphics[width=0.6\textwidth]{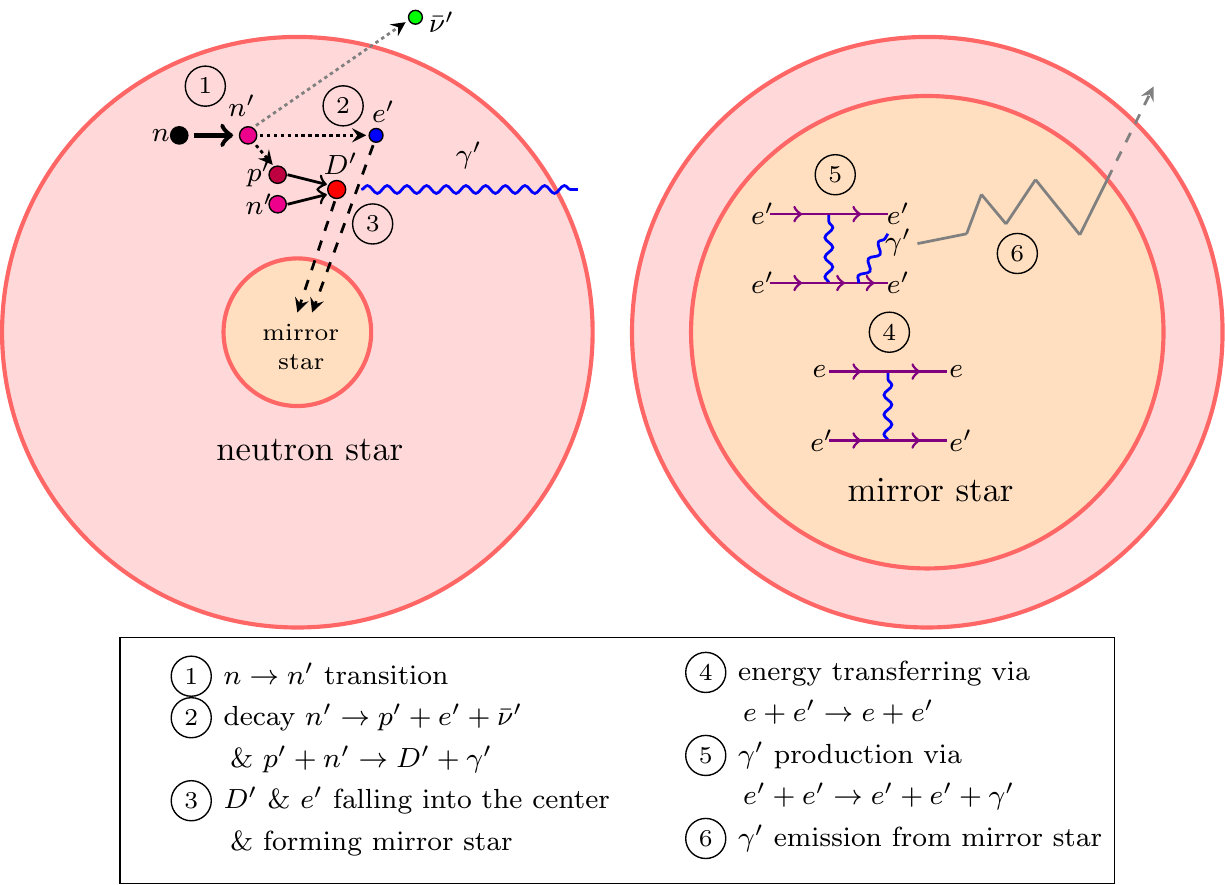}
 \caption{A schematic depiction of what happens after the $n\to n'$ transition takes place in a NS.  In the right panel we zoom in the ``mirror star'' region in the left panel. }
 \label{fig:1}
\end{figure*}

The $e'$ and $D'$ constitute a fluid that is supported against the gravity of the ordinary NS by degenerate pressure, which is dominated by that of the $e'$. The mass density of the fluid is dominated by the $D'$.
The corresponding hydrostatic equation is
\begin{eqnarray}
\label{eqn:hydrostatic}
\frac{\partial}{\partial r}P_{e'}(r) = - \rho(r) g (r) \,,
\end{eqnarray}
where $\rho(r)= n_{e'} (r)m_{D'}$ is the mass density of the $D'$, and the $e'$ pressure for a given Fermi momentum $p_F$ is
\begin{eqnarray}
 P_{e'} = \frac{8\pi}{3 m_e \hbar^3}\int_0^{p_F}dp \frac{p^4}{\sqrt{1 +({p/m_{e'} c)^2} }} \,.
\end{eqnarray}
For the small radii considered, the gravitational acceleration can be approximated by
\begin{eqnarray}
\label{eqn:g(r)}
g(r) = \frac{G_N M(r)}{r^2} = \frac{4\pi}{3}G_N \rho_0 r \,,
\end{eqnarray}
where $G_N$ is the Newtonian constant of gravitation. For $r< 2$ km the density $\rho_0 \simeq 10^{15} \, {\rm gr}\, {\rm cm}^{-3}$ in the center of the NS is almost a constant. The general relativistic modifications of the hydrostatic equation are  very small, at the level of $10^{-3}$. We can solve the hydrostatic equation~(\ref{eqn:hydrostatic}) analytically and get
\begin{eqnarray}
\label{eqn:ne(r)}
n_{e'}(r)= \frac{8\pi}{3m_{e'}^3 c^3 \hbar^3} \left[ \left( \sqrt{X_F^2(0)+1} -  \frac{r^2}{2r_0^2} \right)^2 - 1 \right]^{3/2} \,,
\end{eqnarray}
where $r_0 = (3m_{e'} c^2/4\pi  G_N \rho_0 m_{D'})^{1/2} \simeq 0.296$ km, and $X_F (0) ={p_F} (0)/{m_{e'} c}$ . Then the number of the $e'$ up to the radius $r$ is
\begin{eqnarray}
{\cal N}_{e'}(r)= \int_0^{r}4\pi n_{e'}(x )x^2 dx \,.
\end{eqnarray}

The fluid is confined inside a sphere with radius $R_c$  so that  $n_{e'}( R_c) =0$. Once $X_F(0)$ is given, $R_c$, $n_{e'}(r)$ and the total number ${\cal N}_{e'} = {\cal N}_{D'}$ are determined by pure numbers and fundamental constants. This  resembles the case  of the Chandrasekhar mass. The dimensionless constant $X_F(0)$ is determined by ${\cal N}_{e'}(R_c)= 5\times 10^{47}$ so that ${\cal N}_{e'} = {\cal N}_{D'}$ is half of the total $n'$ generated. We obtain $X_F(0) \simeq 8.9$, implying  $E_F(0) \simeq 4$ MeV and $R_c \simeq 1.18$ km. More details can be found in the supplemental material~\cite{supplemental}.

{\bf  Energy drain from the visible sector to the mirror fluid:--}
In deriving the strict bound by using
the electromagnetic luminosity ${\cal L} = dW/{dt}$ of the NS, a key point is that the rate of $n\to n'$ transition is constant and independent of any thermal or other variations (except for stopping when the mixed star forms, which happens after many Hubble times for the small values of $\epsilon_{nn'}$ considered). The $\sim 50\%$  of the heat generated which
resides in the SM component is then radiated via a fixed black body luminosity~\cite{posp}.
Having all the mirror particles segregated in a ``core region'' {(the orange region in Fig.~\ref{fig:1})} comprising $\sim 0.1\%$ of the star
volume would have seemed  to minimize  their ability to intercept and impede ordinary heat emission and photon
radiation from the mirror free, large outer region. 
This, in turn, would have suggested only minor luminosity reduction and no relaxing of the bounds on $\epsilon_{nn'}$. However, a more careful scrutiny shows that this simplistic argument is misleading.

The energy emission from the core will be dominated by the radiation of mirror photons, while the heat is continuously transferred from the normal sector to the mirror sector by scatterings of the normal and mirror electrons in the core region. For sufficiently large millicharge $\epsilon$, the heat emission rate from the mirror particles
may overtake the normal emission rate from the external surface by an appreciable factor. The ordinary photonic energy may then account only for a small part of the energy generated inside the star. Furthermore, the cumulative effect of this over most of the star's
history will reduce its heat content and push the internal and external surface temperatures to zero, quenching the photonic emission and destroying the steady state model envisioned.

Thanks to the mutual mirror electromagnetic scattering of the mirror particles inside the core region and attendant emission of the fast escaping mirror photons, the time required for their cooling off
and equilibrating at a temperature $T'$ is very short on typical thermal timescale of  $t_{\rm thermal} = W^*/(dW/{dt})$, where $W^*=Q^*$ is the total heat content of the star.
Using Eq.~(\ref{eqn:Qstar}) we find $t_{\rm thermal} \sim 3\times10^{15}$ sec,  which happens to be close to the age of the star.

Since the emission of heat from the mirror sector is much faster than heat transfer between the sectors, any amount of heat in the mirror sector will be emitted rather than go back to the normal sector, which also implies that $T' \leq T$.
To avoid detailed discussion at the particle scattering level, we first view the core region as a black body for the {\it mirror} photons with temperature $T'$,  as indeed it absorbs any such photon falling on it .  The surface of area $4\pi R_c^2$ of the inner ``core region'' serves effectively as an additional boundary, through which the heat in the normal component of the surrounding star can be emitted.
The mirror electrons in the core will then radiate their heat content to the outside with the
rate of black body luminosity: ${\cal L}' = 4\pi \sigma_{\rm SB} R_c^2 T'^4$.
Relative to the internal core region surface $4\pi R_c^2$, the stellar surface is larger --  by roughly a factor of  100. However, the thermal blanket makes the internal temperature about hundred-fold bigger than the surface temperature.
Thanks to the possibility that  $T^{'4} \geq 10^8 T_s^4$, even if we keep $T' < T$ to make $e\to e'$ energy transfers more than the reverse transfer, we can still, in principle, have the rate of mirror photon emission almost six orders of magnitude bigger than that of the
ordinary photons, so long as  $R_c\geq 1$ km.

However, to verify that this indeed happens, we need to check how many $e-e'$ collisions occur per second (which we denote by $\dot{\cal N}_{\rm col}$)
between the ${\cal N}_{e'}( r<R_c)\sim  10^{38}R^3_c \ {\rm cm}^{-3}$ electrons in the core region and the  ${\cal N}_{e'} \sim 5 \times 10^{47}$ mirror electrons. If the total energy transferred per second via these collisions from the ordinary to mirror electrons much exceeds the stellar luminosity, namely the inequality
\begin{eqnarray}
\label{eqn:inequality}
\dot {\cal   N}_{\rm col} \Delta E \sim \dot{\cal N}_{\rm col} \Delta T \gg {\cal L}_{\rm NS} \sim 2\times 10^{36} ~ {\rm keV}~{\rm sec}^{-1}
\end{eqnarray}
holds, then the mirror luminosity dominates and the scenario envisioned in deriving the strict upper bounds on $\epsilon_{nn'}$ becomes inoperative. On the other hand, if
the inequality in Eq~(\ref{eqn:inequality})  is (strongly) reversed, then the scenario above involving the $\beta$ decay of the mirror neutron will be irrelevant.

For the average energy transfer of $\Delta T \sim 0.35$ keV, Eq.~(\ref{eqn:inequality}) becomes $\dot {\cal N}_{\rm col} \geq10^{37}$ sec$^{-1}$. Each electron and also each mirror electron move with the speed of light $c$. Then we can express $\dot{\cal N}_{\rm cal}$ with energy transfer of $\sim$0.35 keV in a manner, which is symmetric between the ordinary and mirror sectors:
\begin{eqnarray}
\dot{\cal N}_{\rm cal} =  \frac{c f f'{\cal N}_e (r< R_c) {\cal N}_{e'} \sigma_{ee'} } {(4\pi/3) R_c^3}  \,,
\end{eqnarray}
where
$f' = kT'/{E'_F} \sim 10^{-4}$ is the fractions of the ``active'' mirror electrons. For $R_c=1.2$ km,
the condition $\dot{\cal N}_c \gg 10^{37}$ sec$^{-1}$ translates into the following requirement on the $e-e'$ scattering cross section:
 \begin{eqnarray}
  \sigma _{ee'} \simeq \epsilon^2 \sigma_{ee} \gg 10^{-50} ~{\rm cm}^{2}  \,,
  \end{eqnarray}
where $\sigma _{ee}$ is the standard Rutherford scattering cross section of electrons in the same kinematic configuration. For the formula for $\sigma_{ee'}$, see the supplemental material~\cite{supplemental}. Including only the Feynman diagram for the $t$-channel photon exchange, the cross section $\sigma_{ee'}$ is calculated by having the relativistic $e$ and $e'$ with energies $E_F \simeq 10 E'_F\simeq 35$  MeV  collide at random relative direction in the laboratory frame and transferring an energy of $T \sim  0.35$ keV between them. Using a plasmon mass as the cutoff, we estimate this cross section to be $\sigma_{ee'} \simeq {4\pi \alpha^2\epsilon^2}/{E_FT} \simeq 10^{-23} \epsilon^2$ cm$^2$,
which leaves us with the rather weak, easy to satisfy requirement
 \begin{eqnarray}
 \label{eqn:epsilon}
  \epsilon^2 \gg 10^{-27}  \,.
 \end{eqnarray}

The strongest upper bounds $\epsilon\leq 10^{-12}$~\cite{Redondo} do not apply here, as in mirror models the dark matter (DM) is made of neutral objects such as the $p'-e'$ composite mirror Hydrogen, deuteron or Helium.  On the other hand, $\epsilon \leq 10^{-9}$ required for cosmological constraint consistent with BBN limits is more directly applicable here~\cite{angela} whereas a weaker limit of  $\epsilon \leq 10^{-7}$ comes from the consistency of asymmetric inflation~\cite{BM}. This still leaves nine orders of magnitude margin for satisfying  Eq.~(\ref{eqn:epsilon}).

We also note that even though the photonic cooling of ultra-cold NS (UCNS) is not a reliable way to set bound on the $n \to n'$ transition rate for near exact mirror  models and slow $n \to n'$ transition, there are situations when it works: e.g. (i) we could have a near exact mirror symmetry but the millicharge of the mirror fermions $\epsilon \leq 10^{-13}$ or, (ii) an asymmetric mirror model with $m_{p'} \geq m_{n'}$ where $n'$ is the DM of the universe, so that $\beta$ decay of mirror neutron is forbidden. It can also work in
other dark baryon contexts, such as those suggested in connection with the neutron
lifetime anomaly.

An advantage of the heating up argument as compared with the orbital period stability method~\cite{GMN} is that:  in the former one can use all pulsars, whereas the latter case requires binary pulsars~\cite{GN, GMN}. Unfortunately, unlike the misquote in Ref.~\cite{posp}, the spinning period changes of single pulsars -- which, as part of the ambitious nano-gravity project,  determined in many cases with stunning accuracy -- {\it cannot} be used, as it is affected by relatively large and incalculable changes due to magnetic braking etc. This is the reason why  binary pulsars were used in Refs.~\cite{GN,GMN}.

{\bf Conclusion.--}
To summarize  our main result is that  the photonic luminosities of UCNSs do not  necessarily imply robust bounds on $\epsilon_{nn'}$. In particular, they do NOT exclude discovery via terrestrial measurements of the tiny $ \epsilon_{nn'}\sim {\cal O}(10^{-17}$ eV). This happens due to the beta decay of $n'$  following {$n\to n'$ transition} {and the subsequent} deuteron formation. Our main assumption, the existence of a millicharge $\epsilon$,  is definitely allowed and possibly even favored within mirror models.
 In this scenario, under the joint effect of the weight of the mirror deuterons and the Fermi energy of the mirror electrons, the mirror deuterons and electrons form a configuration resembling that of a ``mini white dwarf'' inside the NS. A remarkable feature of this configuration is its universality stemming from, and in analogy with, the features of NSs and actual white dwarfs. Within this structure, heat is transferred relatively fast (on characteristic thermal time scales of the NS) from the heat
reservoir in the normal matter of the NS to the mirror sector, and is radiated via mirror photons.

\acknowledgments

{\bf Acknowledgements.--}
The work of R.N.M. is supported by the US National Science Foundation grant no. PHY- 1914631. Y.Z. is supported by the National Natural Science Foundation of China under grant No.\  12175039, the 2021 Jiangsu Shuangchuang (Mass Innovation and Entrepreneurship) Talent Program No.\ JSSCBS20210144, and the ``Fundamental Research Funds for the Central Universities''.


\setcounter{equation}{0}
\setcounter{figure}{0}
\setcounter{table}{0}
\makeatletter
\renewcommand{\theequation}{S\arabic{equation}}
\renewcommand{\thefigure}{S\arabic{figure}}
\renewcommand{\bibnumfmt}[1]{[S#1]}
\renewcommand{\citenumfont}[1]{S#1}

\begin{widetext}
\begin{center}
\textbf{\large Supplemental Material} \vspace{10pt} \\
\textbf{
Itzhak Goldman, Rabindra N. Mohapatra, Shmuel Nussinov and Yongchao Zhang}
\end{center}
\end{widetext}

\maketitle

\setcounter{equation}{0}
\setcounter{figure}{0}
\setcounter{table}{0}
\makeatletter
\renewcommand{\theequation}{S\arabic{equation}}
\renewcommand{\thefigure}{S\arabic{figure}}
\renewcommand{\bibnumfmt}[1]{[S#1]}
\renewcommand{\citenumfont}[1]{S#1}

\section{Solving the hydrostatic equation analytically}

\begin{figure}[!b]
   \centerline{\includegraphics[scale=0.6]{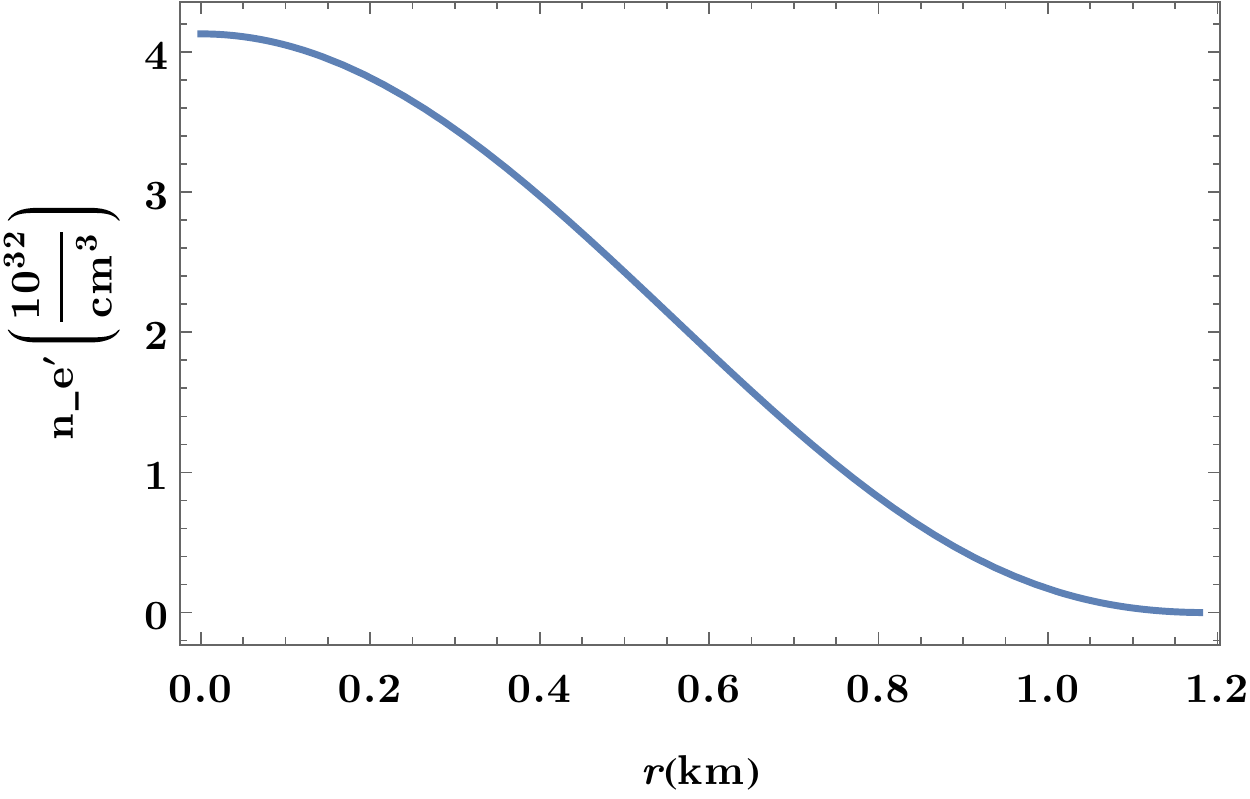}}
 \caption{$n_{e'} $ as function of $r$.}
 \label{fig:2}
\end{figure}

From Eqs.~(\ref{eqn:hydrostatic}) to (\ref{eqn:g(r)}), we obtain
\begin{eqnarray}
 && \frac{8\pi}{3 m_{e'} \hbar^3}\frac{p_F^4}{\sqrt{1 +({{p_F}/{m_{e'} c})^2}}}  \frac{\partial}{\partial r}p_F(r) \nonumber \\
 &=&
- \frac{4\pi}{3}G_N \rho_0  m_{D'}  n_{e'} (r)  r \,.
\end{eqnarray}
Substituting in 
\begin{eqnarray}
\label{eqn:ne_r}
n_{e'}(r)= \frac{8\pi}{ 3}\left(\frac{p_F(r)}{\hbar}\right)^3 \,,
\end{eqnarray}
and introducing the  dimensionless quantity $X_F (r) \equiv {p_F} (r)/{m_{e'} c}$, we arrive at a very simple equation for $X_F(r)$:
\begin{eqnarray}
 \frac{X_F}{ \sqrt{1+ X_F^2} } \frac{d}{dr}X_F(r)
= - \frac{r}{r_0^2} \,.
\end{eqnarray}
The solution is
\begin{eqnarray}
\sqrt{  1+X^2_F(0) } - \sqrt{ 1+X^2_F(r) } = \frac{r^2}{2 r_0^2} \,.
\end{eqnarray}
Then from Eq.~(\ref{eqn:ne_r}) we can obtain the solution for $n_{e'}(r)$ in Eq.~(\ref{eqn:ne(r)}), which is shown in Fig.~\ref{fig:2} as function of $r$.

\section{$e - e'$ scattering cross section}


In this section we give the formulae for $e-e'$ scattering cross section. The scattering of $e$ and $e'$ is very similar to the $e^- - e^-$ M\o ller scattering, with the $e-e'$ scattering having only the $t$ channel diagram, since $e$ and $e'$ are not identical particles. The amplitude square for $e-e'$ scattering is given by
\begin{eqnarray}
\frac14 |{\cal M}|^2 = \frac{2e^2 e^{\prime2} \epsilon^2}{t^2}
\Big[ s^2 + u^2 -8m^2(s+u) + 24m^4 \Big] \,,
\end{eqnarray}
with $m$ the mass of $e$ and $e'$. In the relativistic limit,
\begin{eqnarray}
\frac14 |{\cal M}|^2 = 32 \pi^2 \alpha^2 \epsilon^2
\left[ 1 - \frac{2}{\sin^2\left(\theta/2\right)} + \frac{2}{\sin^4\left(\theta/2\right)} \right] \,,
\end{eqnarray}
with $\theta$ the scattering angle in the center-of-mass frame. Then the differential cross section reads
\begin{eqnarray}
\frac{d\sigma}{d\Omega} &=& \frac{1}{4E_1 E_2} \frac{1}{32\pi^2} \frac14 |{\cal M}|^2 \nonumber \\
&=& \frac{\alpha^2 \epsilon^2}{4E_1 E_2} \left[ 1 - \frac{2}{\sin^2\left(\theta/2\right)} + \frac{2}{\sin^4\left(\theta/2\right)} \right] \,,
\end{eqnarray}
where $E_{1,\,2}$ are the energies of $e$ and $e'$ in the initial state in the star frame.

The presence of $\sin(\theta/2)$ in the denominator implies a mostly forward scattering. The expression is divergent for $\theta=0$ and we put the cutoff at the plasmon mass
in the fluid in our estimate.
In the dense $e-p$ fluid the (ordinary) photon behave as a plasmon with a mass $m_\gamma$ equal to the plasma frequency $\omega$, i.e. $m_\gamma=\omega$. The $e-e'$ scattering cross section is therefore  proportional to $m_\gamma^{-2} = \omega^{-2}$ instead of $(TT')^{-1} \sim T^{-2}$. In normal metals  with $n\sim10^{24}/{\rm cm^3}$, the standard expression
\begin{eqnarray}
\label{eqn:omega}
\omega^2 = 4\pi e^2 n_e/m_e
\end{eqnarray}
yields a plasma frequency corresponding to an energy of $\sim$15 eV. Having here $n_e \sim 10^{38} \; {\rm cm}^{-3}$, i.e. $10^{13}$ times higher, leads to a modification of $e-e'$ cross section, which is $2\times10^{-11}$ times smaller. This dramatically reduces the range of $\epsilon$, for which the basic constraint of ${\cal N}_{\rm col} > 10^{37}$ is satisfied.

However, the plasma frequency in Eq.~(\ref{eqn:omega}) is invalid here. As mentioned in the main text, in the highly degenerate electron fluid only a small fraction $f=T/E_F$ of ``active'' electrons can respond to an external oscillating electric field, much like the fact that only the electrons at the top of the conduction band in metals can freely respond.  In the NS, the electron density $n_e \sim 4k_F^3/{9\pi}$. With the high Fermi energy $E_F \sim 20\ {\rm MeV} \gg m_e$ of the electrons in the NS, the electrons are relativistic, and we can write the density of the relevant ``active'' electrons as:
\begin{equation}
    n_{e,\,{\rm active}} = f \times \frac{4E_F^3}{9\pi} = \frac{4 E_F^2 T}{9\pi} \,.
\end{equation}
Also, the electron mass $m_e$ representing the inertial resistance of the system to oscillating is no longer relevant, and should be replaced by its Fermi enegy $E_F$ in Eq.~(\ref{eqn:omega}). Making these two changes in Eq.~(\ref{eqn:omega}), we have
\begin{equation}
    m_{\gamma}^2 = \omega^2 = \frac{16}{9} E_F T \,.
\end{equation}
The $e-e'$ Rutherford cross section will then be reduced by
\begin{equation}
    \frac{T}{E_F} \sim \frac{0.35 \rm \; keV}{20 \rm \; MeV} \sim 10^{-5} \,,
\end{equation}
which still allows ${\cal N}_{\rm col} > 10^{37}$ so long as $\epsilon > 10^{-13}$.

\end{document}